# LDSF: Lightweight Dual-Stream Framework for SAR Target Recognition by Coupling Local Electromagnetic Scattering Features and Global Visual Features

Xuying Xiong[a], Xinyu Zhang[a,*], Weidong Jiang[a], Tianpeng Liu[a]

*a. College of Electronic Science and Technology, National University of Defense Technology, Changsha 410073, China*
*b.*

A B S T R A C T

SAR-ATR research based on deep neural networks (DNNs) has made many notable achievements in recent years. However, mainstream DNN-based methods still face issues such as easy overfitting of a few training data, high computational overhead, and poor interpretability of the black-box model. Integrating physical knowledge into DNNs to improve performance and achieve a higher level of physical interpretability becomes the key to solving the above problems. This paper begins by focusing on the electromagnetic (EM) backscattering mechanism. We extract the EM scattering (EMS) information from the complex SAR data and integrate the physical properties of the target into the network through a dual-stream framework to guide the network to learn physically meaningful and discriminative features. Specifically, one stream is the local EMS feature (LEMSF) extraction net. It is a heterogeneous graph neural network (GNN) guided by a multi-level multi-head attention mechanism. LEMSF uses the EMS information to obtain topological structure features and high-level physical semantic features. The other stream is a CNN-based global visual features (GVF) extraction net that captures the visual features of SAR pictures from the image domain. After obtaining the two-stream features, a feature fusion subnetwork is proposed to adaptively learn the fusion strategy. Thus, the two-stream features can maximize the performance. Furthermore, the loss function is designed based on the graph distance measure to promote intra-class aggregation. We discard overly complex design ideas and effectively control the model size while maintaining algorithm performance. Finally, to better validate the performance and generalizability of the algorithms, two more rigorous evaluation protocols, namely once-for-all (OFA) and less-for-more (LFM), are used to verify the superiority of the proposed algorithm on the MSTAR.

## 1. Introduction

Synthetic Aperture Radar (SAR) is an active microwave remote sensing imaging radar(Moreira et al., 2013), it provides all-day and all-weather image acquisition capabilities. In recent years, SAR data has become increasingly abundant, and spatial resolution has become higher(Castelletti et al., 2021). SAR data is gradually meeting the demand for precise resolution and interpretation of targets/objects, playing an important role in multiple fields such as target reconnaissance and monitoring, strike indication, and natural disaster response(XU et al., 2020). Automatic target recognition (ATR) refers to the detection and recognition of target features and models from SAR images.

With the significant progress made in computer vision (CV), machine learning based methods have gradually become mainstream in SAR-ATR. CNN-based methods are the typical representative.(Chen et al., 2016; Geng et al., 2015; Lin et al., 2017). The application data in CV mainly comes from optical sensors, although many impressive progresses have been achieved by applying the experience gained from processing optical images to SAR-ATR.

However, from a practical perspective, the current mainstream methods still face some common challenges. First of all, the small volume of SAR data (relative to optical images) makes it extremely easy to cause model overfitting. Conventional solutions to the few-shot learning problem include data extension and model optimization(Y. Li et al., 2022; Ren et al., 2023; Wang et al., 2022; Y. Zhao et al., 2023). But both of these will lead to an increase in computational overhead and an increase in the parameter of the model. At the same time, the algorithms will be difficult to deploy to real engineering application endpoints. Secondly, the inherent characteristics of radar and the differences between radar images and optical images bring difficulty in SAR-ATR. Visual angle sensitivity is an important characteristic of SAR, which makes SAR have significant intra-class differences and high inter-class similarity(Luo et al., 2023). This characteristic can easily cause feature space confusion. In addition, the practice of directly feeding SAR images into the network actually utilizes only the magnitude of the SAR(Feng et al., 2022). However, most of the rich information obtained from radar sensors is embedded in the phase (J. Liu et al., 2022; Zhang et al., 2017). Thus, using only the magnitude image of the SAR has already lost a considerable amount of information at the input. Finally, the black box characteristic of DNNs leads to lower credibility of DNN among users and limits decision-making ability in some task scenarios (Huang et al., 2021).

---

\* Corresponding author.
*E-mail address*: zhangxinyu90111@163.com (X. Zhang), xiongxuying@nudt.edu.cn (X. Xiong)

The research on deep learning applications in SAR-ATR has been advanced quite deeply. Continuing the classic data-driven network improvement approach may achieve some indicator improvements on a specific dataset, but it still cannot fundamentally overcome the above problems. To meet these challenges, the key is to develop DNNs that differ from pure data-driven methods(Karniadakis et al., 2021). SAR physical characteristics reflect the radar imaging mechanism and the intrinsic characteristics of the target. It is an important basis for researchers to carry out SAR-ATR(HUANG, 2005). Recently, there have been many top research teams focusing on the importance of radar characteristics for SAR-ATR. They are trying to transform domain knowledge and physical models into the network structure to solve the problems of SAR-ATR while enhancing the interpretability of models(Datcu et al., 2023; Huang et al., 2022a; Karpatne et al., 2017; Liao et al., 2022). These improved approaches have two benefits. One is that the network can not only mine the laws from the data but also learn from the physical laws summarized by human beings. Therefore, the network's requirement for the amount of training data can be reduced to a certain extent(Liu et al., 2023). On the other hand, a priori knowledge based on physical models (e.g. SAR imaging mechanism, physical scattering properties of the target, magnitude and phase information, etc.) is considered to have clear interpretability(Datcu et al., 2023; Huang et al., 2022a). Using physical knowledge to bootstrap DNNs can enable networks to learn high-level semantic features that have physical awareness, and enhance the physical interpretability of the network and the credibility of the results(Huang et al., 2022b). A lightweight dual-stream framework (LDSF) constructed in this paper effectively combines physical knowledge and fully exploits the target characteristics, which can realize a better and stabler performance. The physical knowledge used in LDSF is EMS information. EMS information can improve the performance and generalizability of the network, due to its invariance to azimuthal variations and the robustness of signal-to-noise ratio (SNR) variations(Ding et al., 2018; Y. Li et al., 2022; Zhang et al., 2021). Specifically, LDSF consists of local EMS features (LEMSF) extraction net and global visual features (GVF) extraction net.

How EMS information is acquired and utilized is critical to improving the performance of the algorithm. We improved the parameter estimation method of the Attribute Scattering Center (ASC) model to construct the EMS information acquisition (EMSIA) module, which obtains physical knowledge from complex SAR data (Gerry et al., 1999; Potter et al., 1995; Potter and Moses, 1997). The scattering centers obtained from the ASC model correspond to the key components of the target. Based on the physical information, LEMSF constructs a multi-level multi-head attention mechanism (MMAM) guided heterogeneous GNN to learn local EMS features. Local EMS features include target topology features, component categories information, and component EM semantic information.

The topological relationship of target components can represent the inherent morphology and configuration of the target such as the rectangular structure of vehicles and the cross structure of airplanes(Luo et al., 2023). According to statistical analysis, the convergence of visual features such as color, size, and orientation is very dependent on the quantity of samples. Nevertheless, the variation of topological features with the number of samples is relatively minimal and smooth, and the extreme values are limited(Zhu et al., 2022). Under few training data conditions, topological features can better represent the target than other features(C. Li et al., 2022; Peng et al., 2022; C. Zhao et al., 2023; Zhu et al., 2023). It can also enhance the model's capacity to discern local structural changes in the target, and alleviate visual angle sensitivity(Zhang et al., 2021). However, the topological features belong to non-Euclidean spatial data. The image and text data processed by CNNs are all in Euclidean space. CNNs are sensitive to changes in rotation, scale, etc., making it difficult to extract topological features(Ding et al., 2019). GNN is a network that can effectively deal with non-Euclidean spatial data (Bronstein et al., 2017; Wu et al., 2021; K. Xu et al., 2019; Z. Zhang et al., 2022; Zhou et al., 2020). GNNs represent any entity and its relationships as a graph, thereby learning features from the graph(Hamilton, 2020; H. Xu et al., 2019). Recently, GNNs have made significant achievements in multiple fields(Fan et al., 2022; Ktena et al., 2017; Liu et al., 2019; Parisot et al., 2017). In the field of remote sensing, some scholars have also started to study the application of GNNs (Kang et al., 2023; C. Li et al., 2022; Li et al., 2023; Zhang et al., 2023; C. Zhao et al., 2023). In addition to the topological relationship of target components, the category information of components is also a kind of semantic information, which is helpful for recognition (Y. Li et al., 2022). We incorporated component category information into LEMSF by constructing heterogeneous graphs. The EM semantic information of components is contained in the graph's node attribute vectors.

Beyond strong reflections in the image, the contours and shadows of the target also contain information that helps in classification and they belong to the visual features. Visual features include both the overall picture of the image and the local visual features of the target. Therefore it is also known as global visual features (Ding et al., 2018; Feng et al., 2022; Y. Li et al., 2022; C. Zhao et al., 2023). GNN lacks the ability to learn these features. Inspired by CNNs' successful application in ATR, another stream of LDSF, global visual features (GVF) extract global features by constructing a CNN network (Ding et al., 2018; Y. Li et al., 2022; Penatti et al., 2015). There is an essential difference between the local visual features and the local EMS features. Even if CNNs use the small-sized kernel to get the local features of the image, they cannot learn the topological relationships and phase information between these local features. Therefore, the main goal of the GVF is to learn intuitive visual features from the image domain of the picture. Moreover, we controlled the volume of network parameters in the design process, ensuring the algorithm's operational speed and engineering deployment ability.

We can obtain feature descriptions about different sides of the target by LEMSF and GVF. Utilizing multiple features will inevitably face the problem of how to effectively perform feature fusion. To this end, we design the feature fusion subnet to adaptively fuse the two features, so that they can maximize the performance of the algorithm. In summary, the main innovations and contributions are as follows:

1. To our knowledge, we are the first to introduce a heterogeneous GNN in SAR-ATR for innovatively integrating topology, component types, and EM semantics information of the target. Furthermore, we devise the MMAM mechanism to facilitate the exploration of intricate interaction characteristics and multi-level semantic richness within the heterogeneous graph by the GNN. Additionally, drawing upon graph distance measure theory, we refine the loss function, thereby bolstering the intra-class aggregation of local EMS features.

2. An improved image-domain ASC parameter estimation algorithm is proposed to more efficiently and accurately extract EMS information from SAR complex data.
3. To ensure the maximum contribution of two stream features to the classification task, while avoiding the negative impact of the poor flexibility of physical models. We considered the distinction and connection of multi-stream features thoroughly and constructed a feature fusion subnet to adaptively integrate two streams of features.
4. On the premise of ensuring model performance, LDSF incorporates effective network pruning techniques during its design process to enhance the engineering deployment ability of the algorithm.

Section 2 provides a review of the cutting-edge work in the related fields and summarizes the ideas of the latest research results. Section 3 provides a detailed description of the specific technical details of LDSF. Section 4 is based on two more rigorous evaluation protocols, OFA and LFM, to verify the superiority of LDSF. Section 5 summarizes all the work.

## 2. Related work

The overall structure of LDSF is shown in Fig. 1, the entire network consists of two streams and four modules. The input of LDSF is the complex raw SAR data. This section focuses on the knowledge and latest research results related to LDSF.

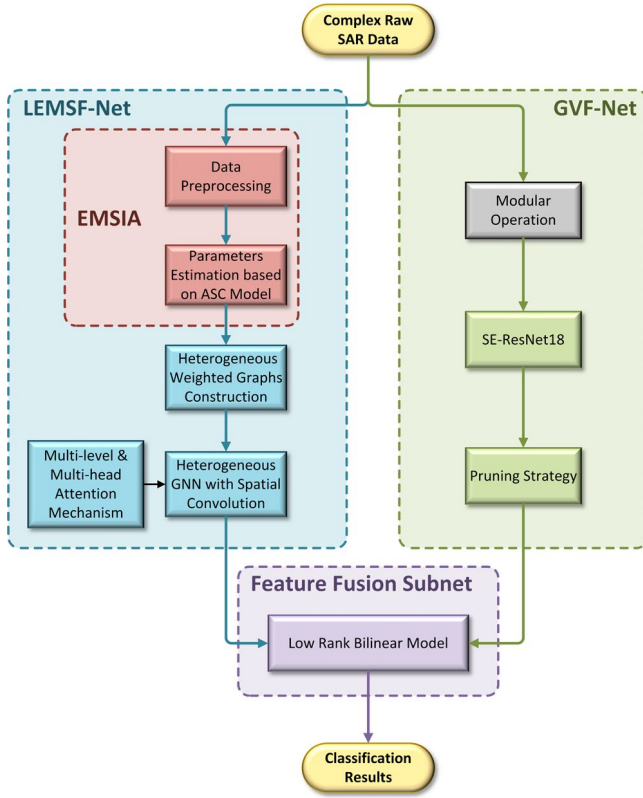

Fig. 1 Overall structure of LDSF

### 2.1. Extraction and Utilization of EMS Information

The physical knowledge for LDSF is the EMS information parsed from ASC parameters. The current ASC parameters estimated algorithms are categorized into two groups: image-domain algorithms(Koets and Moses, 1999) and frequency-domain algorithms(Cong et al., 2016; Li et al., 2016; Yang et al., 2020). Frequency-domain algorithms require joint estimation of all parameters with high arithmetic complexity and need to predetermine the order of the model. Image-domain algorithms have no restriction on the model order and have faster computational efficiency (Yang et al., 2020). Thus, we design an image-domain algorithm to construct EMSIA. There are two key steps in ASC parameters estimation: scattering center extraction and optimization function calculation. We have added a distance constraint to the scattering center extraction step, which effectively avoids mis-extracting background points as target points. BFGS Quasi-Newton and Nelder-Mead are used to calculate the optimization function, which can effectively obtain more accurate ASC parameters.

Recently, numerous studies have highlighted the significance of EMS information in SAR-ATR applications. Some work has transformed EMS information into bags of words or feature vectors, subsequently splicing vectors with deep features derived from the network(Z. Liu et al., 2022; Zhang et al., 2021). (Feng et al., 2023, 2022) and (Y. Li et al., 2022) utilized EMS information to reconstruct the images of typical components as inputs to the network. (J. Liu et al., 2022) employed EMS information to develop convolution kernels for the initial convolutional layer within CNN architectures. Nonetheless, the above post-processing techniques typically forfeit the original information of the scattering center, extremely relying on transformed semantic data(Zhang et al., 2023). Furthermore, they neglected the correlations among scattering centers and failed to exploit structural information. (Kang et al., 2023; C. Li et al., 2022; C. Zhao et al., 2023) concentrated on the spatial topological relationships among scattering centers. However, these approaches only utilized the coordinates and amplitude of scattering centers while neglecting the inherent physical properties.

The method by which physical information is employed heavily impacts the effectiveness of input data to the network. We obtain physically meaningful EMS information by simultaneously considering the ASC parameters as well as the topological relationship of scattering centers. This approach helps us avoid any loss of data due to excessive post-processing and enables the network to use physical knowledge effectively.

### 2.2. Learning Local EMS Features through GNN

Graphs are a mathematical tool used to represent the relationships between different entities. For a typical graph $\mathbf{G}=(\mathbf{V},\mathbf{E})$. $\mathbf{V}=\{v_1, v_2, \cdots, v_m\}$ represents the nodes of the graph, and $\mathbf{V} \in \mathbb{R}^{|\mathbf{V}| \times 1}$, $|\mathbf{V}|$ is the quantity of nodes. $\mathbf{E}=\{e_1, e_2, \cdots, e_n\}$ represents the edges. The adjacency matrix $\mathbf{A}$ is the topological representation of $\mathbf{G}$, and $\mathbf{A} \in \mathbb{R}^{|\mathbf{V}| \times |\mathbf{V}|}$. For a weighted graph, if node $v_i$ and node $v_j$ have a connection, $a_{ij}$ is equal to the weight value $\omega$ otherwise equal to 0.

Building a graph in this space has more benefits than CNNs for learning topological features (Battaglia et al., 2018; Zhu et al., 2022). The scattering centers can be visualized as nodes on a graph, while the relationship between them can be denoted as edges. Despite awareness among certain researchers that incorporating topological features can enhance classification performance(Peng et al., 2022). There is a lack of applications of GNNs in the SAR domain and current methods face significant limitations.

#### 2.2.1. Node construction

Three key issues need to be addressed when constructing graph nodes. **The first is the construction of node attribute vectors.** In (Kang et al., 2023), graph nodes were derived by applying the Harris operator on amplitude images. After that, the L-neighborhood features around each coordinate were extracted from the corresponding position on the output feature map of the first CNN layer, and used as the attributes of the nodes. However, this approach fails to capture the inherent physical properties of the nodes adequately. The node attributes in (C. Li et al., 2022; C. Zhao et al., 2023) only included the amplitude and position information of the scattering centers. In (Li et al., 2023), the reconstructed scattering center images based on ASC parameters were multiplied by the feature map obtained from the first CNN convolution of the original image. Then the feature vector after global pooling of the multiplication result was used as the attributes of the nodes. However, the above methods tend to overlook the EMS information of the scattering centers themselves(Zhang et al., 2023).

**The second is the determination of the number of nodes.** It is common to extract scattering centers with a pre-set fixed number (C. Li et al., 2022; Li et al., 2023; C. Zhao et al., 2023). However, due to the visual angle sensitivity, even for the same category of targets, the quantity of scattering centers may differ across various visual angles. Furthermore, if the actual number of scattering centers falls short of the pre-set count, there's a high probability that background points will be mis-extracted as target points. Zhang et al. (2023) developed two methods to extract a large number of nodes and then remove unreasonable nodes. Although this approach ensured a thorough extraction of the key components, it significantly increased the computational load. Kang et al. (2023) also obtained nodes in two steps but ultimately retained only 9 nodes to represent components at different locations of the aircraft. However, the points extracted from the image domain lack distinct physical information, making it difficult to ensure that the nodes indeed correspond to key components of the target. The effectiveness of the node filtering strategy would directly affect the quality of the graph in these two-step approaches.

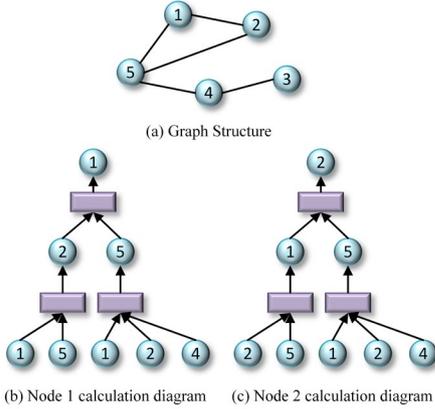

Fig. 2 Node 1 and Node 2 have the same computational process. In information aggregation, homogenous graphs cannot distinguish between the two processes.

**Thirdly, the types of nodes need to be distinguished.** All previous studies considered all scattering centers as the same type, which means the graphs they constructed are all homogeneous(Kang et al., 2023; C. Li et al., 2022; Li et al., 2023; Zhang et al., 2023; C. Zhao et al., 2023). However, from the perspective of GNN's computational principle, nodes in homogeneous graphs may not be distinguishable if node attribute information is lacking. For instance, node 1 and node 2 in Fig. 2 undergo the same computational process, making them isomorphic in the case of a homogeneous graph (Dong et al., 2017; K. Xu et al., 2019). Therefore, the homogeneous graph requires higher accuracy for node attribute vectors.

To meet the above challenges, we classify scattering center categories to build heterogeneous graphs. This approach enables us to incorporate more semantic information and better reflect the nature of the target. Heterogeneous graphs are commonly used in the real world (Hamilton, 2020; Shi et al., 2017). In comparison to homogeneous graphs, heterogeneous graphs can include more domain knowledge in the learning process of GNNs (Shi et al., 2017). Our work constructs a heterogeneous graph with multiple node types, where the node types are categorized based on the EMS information. We use the ASC parameters to build the attribute vectors of the nodes. Additionally, we do not follow the practice of fixing the number of nodes in the graph. Instead, we extract the key nodes of the target by analyzing the energy proportion of the image.

*2.2.2. Edge construction*

The connection relationship of nodes is represented by the edge. Existing edge construction methods have certain limitations. **First of all, the threshold determination criterion may introduce additional errors.** Most edge construction methods first calculate the actual distance between nodes, and then set a threshold. When the distance is shorter than the threshold, an edge is constructed (Kang et al., 2023; C. Li et al., 2022; Li et al., 2023; C. Zhao et al., 2023). However, the setting of the threshold lacks scientific guidance, and the value of the threshold affects the classification effect directly(C. Zhao et al., 2023). Different categories of targets may not necessarily apply to one threshold. Therefore, an inappropriate threshold will add additional errors during graph construction, which will reduce classification performance.

**Secondly, the similarity premise may not necessarily hold.** Another widespread edge-constructing method is to devise a measurement function to gauge the similarity among nodes. The commonly used similarity criterion is cosine similarity. Specifically, when the cosine similarity between nodes is bigger than 0, then the edge is constructed (Zhang et al., 2023). However, this premise may not necessarily hold in the SAR-ATR. For example, the similarity between a tank's gun barrel and its tracks is extremely low, but the positional relationship between them is very important. Therefore, using nodes' similarity as the principle of edge construction lacks strong theoretical support in the SAR-ATR, and there is a risk of incorrect edge construction.

**Thirdly, most of the current methods construct unweighted graphs**. The weight of edges can describe the influence between nodes (Kang et al., 2023). Therefore, assigning weights to edges can further increase the information of the graph. We use the actual distance between nodes to construct edge weight. However, the coordinates in ASC parameters are on the slant range plane. Thus, the coordinates of nodes should be converted before edge weights canulation.

*2.2.3. Heterogeneous GNN with spatial convolution*

According to the previous analysis, the graphs constructed in LEMSF are heterogeneous graphs with diverse node types. The convolutional criterion on homogeneous graphs is no longer applicable in heterogeneous graphs(Kang et al., 2023; Li et al., 2023; Zhang et al., 2023). Therefore, LEMSF uses type-related transformation matrices to project the attribute vectors of different types of nodes into a common implicit space. Additionally, the number of nodes in our graph is not fixed, since we extract nodes based on the energy proportion of the image. For this reason, we construct the GNN architecture under spatial domain convolution based on the message-passing mechanism (Chen et al., 2023). Considering the variation in contribution between nodes and node types, MMAM is introduced into LEMSF to effectively capture key information from all levels (Yang et al., 2021). The specific technology details will be discussed in Section 3.2.3. For the problem of large intra-class variation in SAR images, we introduce the graph distance measure into the loss function to improve intra-class aggregation (Backhausz and Szegedy, 2022; Borgs et al., 2012, 2008; Lovász, 2012).

## 2.3. Learning visual features through CNN

LEMSF has a limitation in effectively utilizing visual features such as target shadows and backgrounds in images. Due to the characterization and outstanding performance of CNNs in visual feature extraction, we have chosen CNN as the backbone of GVF to extract critical visual features.

It is important to note that the scale of LEMSF is relatively small, and we intend to maintain that advantage. With the help of LEMSF's capability to efficiently extract local EMS features, the branch responsible for extracting visual features does not require overly complicated network layers and parameter designs. Our design philosophy is to expect the GVF to have the appropriate ability to extract global visual features, rather than blindly pursuing greater depth. Therefore, we utilize a pruning strategy to balance the network scale in GVF design.

Based on the application experience of CNN models in the SAR-ATR task, we have selected SE-ResNet18 (Hu et al., 2020) as the baseline of GVF. The specific technology details will be introduced in Section 3.2.4.

## 2.4. Features Fusion Strategy

LEMSF learns the physical features of complex data, while GVF learns the visual features of amplitude images. Z. Liu et al. (2022) found that the fusion method of the features has a significant impact on the final classification effect. Due to the differences in information and semantic levels between the two features(Huang et al., 2022b; Zhang et al., 2021), it is important to find a way to fuse and utilize features reasonably to achieve better classification results (J. Liu et al., 2022; Xing et al., 2022).

Since the feature set contains richer information than the matching score or the output decision of the classifier, fusing at the feature level is considered an effective fusion method (Haghighat et al., 2016). The previous work has approached the two types of features differently at the feature layer and has found that the optimal fusion strategy varies across different task scenarios (Y. Li et al., 2022; Z. Liu et al., 2022; Zhang et al., 2021). We believe that using network-adaptive learning feature fusion strategies is more scientific than manually designing fusion methods and fusion weights (J. Zhang et al., 2022). Therefore, we construct a feature fusion subnet to ensure that the two types of features are combined to optimally support the LDSF classification.

## 3. Methodology

### 3.1. EMS Information Acquisition (EMSIA)

LDSF extracting EMS information based on the ASC model (Gerry et al., 1999; Potter et al., 1995; Potter and Moses, 1997). ASC model describes the complex EMS characteristics of targets in high-frequency regions. The mathematical form of the ASC model is shown in Eq.(1)

$$E(f,\phi;\Theta) = \sum_{i=1}^{P} E_i(f,\phi;\theta_i) \quad (1)$$

Eq.(1) indicates that when the azimuth angle is equal to $\phi$ and the frequency is equal to $f$, the target overall backscattering is composed of $P$ independent scattering centers. It should be noted that $P$ changes with the variations of $\phi$. $\Theta$ represents the sets of parameters for all scattering centers, i.e., $\Theta = \{\theta_i\}(i=1,2,\ldots,P)$, where $\theta_i$ is the parameter vector of the $i$th scattering center. $E_i(f,\phi;\theta_i)$ represents the independent response of the $i$th scattering center, the specific form is shown in Eq.(2)

$$\begin{aligned}E_i(f,\phi;\theta_i) = &A_i\left(j\frac{f}{f_c}\right)^{\alpha_i} \\ &\cdot \text{sinc}\left(\frac{2\pi f}{c}L_i\sin(\phi-\varphi_i)\right) \\ &\cdot \exp\left(-j\frac{4\pi f}{c}(x_i\cos\phi+y_i\sin\phi)\right) \\ &\cdot \exp(-2\pi f\gamma_i\sin\phi)\end{aligned} \quad (2)$$

where the center frequency of radar is denoted as $f_c$, $c$ represents the speed of light. We can get the specific form of $\theta_i$ from Eq.(2), i.e., $\theta_i = \{A_i,\alpha_i,L_i,\varphi_i,\gamma_i,x_i,y_i\}$. Obtaining the ASC parameters allows us to reveal the EMS information of the target components as well as the topology relationship(Gerry et al., 1999). To estimate ASC parameters, we need to find the optimal vector that minimizes the difference between the reconstructed and actual images. The mathematical form is

$$\hat{\Theta} = \arg\min_{\theta}\left\|\mathcal{F}^{-1}\{E(f,\phi)\} - \mathcal{F}^{-1}\left\{\sum_{i}^{P}E_i(f,\phi;\theta_i)\right\}\right\|^2 \quad (3)$$

where $\mathcal{F}^{-1}\{\cdot\}$ represents inverse Fourier transform, $E(f,\phi)$ represents actual collected data. The ASC parameters estimation process in EMSIA is shown in Table 1.

**Table 1**
ASC parameters estimation algorithm pseudocode

| **Algorithm**: ASC parameters estimation based on regional decoupling | |
|---|---|
| **Input:** | $E(f,\phi)$;<br>Algorithm termination condition |
| **Output:** | Scattering centres' parameters: $\hat{\Theta} = \{\theta_j\}(j=1,2,\ldots,J)$ |
| 1 | Use the 2D OTSU algorithm(Otsu, 1979) to obtain target area $E_{seg}$ and $I_{seg} = \mathcal{F}^{-1}\{E_{seg}(f,\phi)\}$;<br>Scattering center count initialization: $j := 1$;<br>Residual image initialization $I_{res} := I_{seg}$. |
| 2 | **while** termination conditions not met **do** : |
| 3 | Use the improved watershed algorithm to segment the region $E_{max}$ from $I_{res}$, and determine the type of scattering center corresponding to $E_{max}$. |
| 4 | Calculate the initial attribute parameter $\theta_{j0}$ corresponding to $E_{max}$ based on the ASC model and SAR image characteristics. |
| 5 | Use BFGS Quasi-Newton and Nelder-Mead respectively to solve Eq.(3) the estimated parameter that makes the value of the cost function smaller as $\theta_j$. |
| 6 | Get reconstructed images: $I_{rebu} = \mathcal{F}^{-1}\{E_j(f,\phi,\theta_j)\}$. |
| 7 | Update residual images: $I_{res} := \mathcal{F}^{-1}\{E_{res}\} - I_{rebu}$. |
| 8 | $j := j+1$ |

ps: The image-domain data is denoted by $I$, and the frequency-domain data is denoted by $E$. Footmarks are used to distinguish the results of different stages in the algorithm.

There are three termination conditions for algorithm iteration. The first is the maximum fitting percentage, which is the proportion of the extracted scattering center's energy in the total energy of the original image. The second is the minimum peak level of scattering centers, defined as the ratio of the final segmented scattering center peak to the highest peak in the original image. The third is the maximum number of scattering centers.

$E_{max}$ is the scattering center region with the highest amplitude in the current region $I_{res}$. To prevent the extraction of strong scattering points or coherent speckle noise in the background, a scattering center judgment is added for the improved watershed algorithm in step 3. Specifically, $E_{max}$ will be re-extracted, if the Euclidean distance between $E_{max}$ and extracted regions is too large. After obtaining $E_{max}$, the type of corresponding scattering center can be distinguished. For local scattering centers, $L_i = \varphi_i = 0$, while for distributed scattering centers, $\gamma_i = 0$. Step 4 adopts the same method as that in (JI, 2003). Step 7 introduces the CLEAN idea (Koets and Moses, 1999) to remove the side lobe effect of estimated scattering centers on the estimated scattering centers. When the iteration terminates, the number of scattering centers is $j$. The extraction results obtained through the above algorithm are shown in Fig. 3.

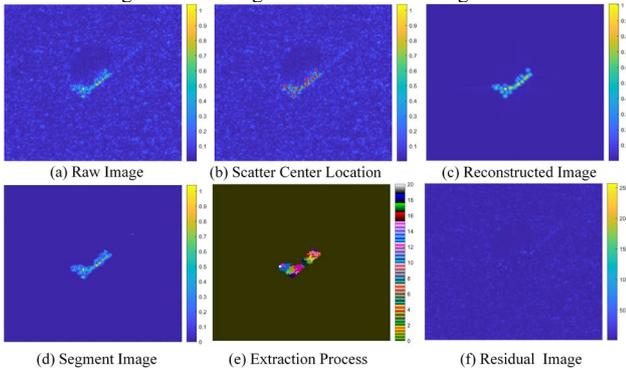

Fig. 3 Scattering center extraction results. (a) is the original image, (b) marks the position of scattering centers with red dots based on the original image, (c) is the result reconstructed based on estimated parameters, (d) is the target segmentation result obtained by using improved 2D-OTSU, (e) uses different colors to represent different scattering center regions, (f) is the residual image remaining after subtracting the estimated target region from the original image.

Fig. 3(d) demonstrates the effectiveness of the target segmentation. Fig. 3(c) shows that the ASC parameters estimated based on EMSIA can reconstruct the various scattering structures of the target in a relatively complete manner. In Fig. 3(f), only the target background clutter and a small amount of edge structure estimation error are visible. These results prove the effectiveness of the ASC model parameters, which can be used as physical information for subsequent target recognition methods.

### 3.2. Local EMS Feature Extraction (LEMSF) Net

#### 3.2.1. Heterogeneous graph construction

We used the ASC parameters to construct heterogeneous graphs to represent target-embedded EMS information. First, local scattering centers and distributed scattering centers are denoted as two types of nodes in the graph, which are denoted as $\mathbf{V}_1$ and $\mathbf{V}_2$, respectively. Then the node set of the entire graph is represented as $\mathbf{V} = \mathbf{V}_1 \cup \mathbf{V}_2 = \{v_1, \ldots, v_i, \ldots, v_m\}$. Huang et al. (2024) experimentally proved that a complete description of the scattering center provides more complete physical information about the SAR target. To avoid loss of EMS information due to excessive processing, the initial attribute vector of nodes $v_i$ consists of the ASC parameters corresponding to the scattering center, i.e., $\mathbf{x}_i = (A_i, \alpha_i, \gamma_i, L_i, \varphi_i, x_i, y_i)$. The node attribute matrix is denoted as $\mathbf{X} \in \mathbb{R}^{|V| \times 7}$.

In cases where the relationship between nodes is unclear, constructing a fully connected graph is an effective way to avoid information omission. The target physical characteristics of SAR determine that the number of nodes in the graph is relatively small, and the number of edges does not change the dimension of $\mathbf{A}$. Therefore, constructing a fully connected graph does not impose too much computational burden. As two nodes are further apart, their correlation becomes weaker. The edge weight shown in Eq.(4) is represented by the reciprocal of the actual distance $d_{ij}$ between the scattering centers.

$$\omega_{ij} = \frac{1}{d_{ij}} \tag{4}$$

However, the coordinates obtained from the ASC model are in the imaging plane. Due to the visual angle sensitivity, it is more reasonable to use the true physical distance between components to calculate edge weight. This approach can be more closely aligned with the structural characteristics of the target itself. As shown in Fig. 4, in SAR image formation, the received backscattered returns are projected into a "slant range plane", which is defined by the radar line-of-sight to the map center and the platform velocity vector. $\phi$ presents squint angle, $\beta$ presents depression angle. Fig. 4(b) shows the projection relationship between the SAR imaging plane and the ground plane.

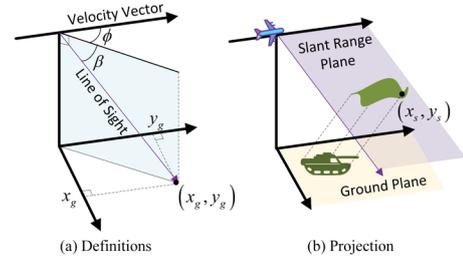

Fig. 4 The projection relationship of coordinates between the SAR imaging plane and the ground object plane.

$(x_s, y_s)$ is the coordinate of the target projected on the imaging plane, $x_s$ represents the distance between the scattering center and the radar platform at the synthetic aperture center moment. $(x_g, y_g)$ is the coordinate of the scattering center in the ground plane. The coordinate conversion rules are as follows:

$$\begin{cases} x_g = x_s \cos\theta \cos\phi \\ y_g = y_s \end{cases} \tag{5}$$

Because the origin of the azimuth coordinate is based on the starting time of data collection, and the ground distance y-axis and imaging plane y-axis are parallel, we have $y_g = y_s$. The ground coordinates of node $v_i$ and node $v_j$ are denoted as $(x_{gi}, y_{gi})$ and $(x_{gj}, y_{gj})$, respectively. The actual distance between two nodes can be calculated from:

$$d_{ij} = \sqrt{(x_{gi} - x_{gj})^2 + (y_{gi} - y_{gj})^2} \tag{6}$$

Then perform symmetric normalization on $\mathbf{A}$:

$$\tilde{\mathbf{A}} = (\mathbf{D} + \mathbf{I})^{-\frac{1}{2}} (\mathbf{I} + \mathbf{A})(\mathbf{D} + \mathbf{I})^{-\frac{1}{2}} \tag{7}$$

where $\mathbf{D}^{|V| \times |V|}$ is the degree matrix, which is a diagonal matrix defined as: $d_{ii} = \sum_{j \in N(v_i)} \omega_{ij}$, where $N(v_i)$ is the set of all nodes which are connected to the node $v_i$. $\mathbf{I}$ is an identity matrix. It is important to note that certain graphs may only contain a single node type, which does not affect the subsequent training of the network.

#### 3.2.2. Heterogeneous graph convolution

The structure of the convolutional layer in homogeneous GNN (Kipf and Welling, 2017) is:

$$\mathbf{H}^{(l+1)} = \sigma\left(\tilde{\mathbf{A}} \cdot \mathbf{H}^{(l)} \cdot \mathbf{W}^{(l)}\right) \quad (8)$$

where $\mathbf{H}^{(l)}$ represents the $l$th output feature, and $\mathbf{H}^{(l)} \in \mathbb{R}^{|V| \times q^{(l)}}$, $q^{(l)}$ denotes the length of the node attribute vector in the $l$th layer. Initially, $\mathbf{H}^{(0)} = \mathbf{X}$. $\mathbf{W}^{(l)}$ represents the trainable weight matrix in the $l$th layer, $\mathbf{W}^{(l)} \in \mathbb{R}^{q^{(l)} \times q^{(l+1)}}$. $\sigma(\cdot)$ is the activation function. Heterogeneity makes the homogenous graph convolution criterion no longer applicable. A direct solution is to orthogonally splice different types of nodes to their respective feature spaces. However, such a method will result in a huge feature space, and it still ignores the differences between different types of nodes, and the actual effect is not ideal (Linmei et al., 2019).

To better express the topological relationships between components in the target and the semantic information derived from them, we extract low-level and high-level features from three perspectives: nodes, semantics, and graphs. Firstly, different node types have different characteristics, and their features fall into different feature spaces. Therefore, we multiply each node type with its corresponding transformation matrix. In this way, different node information can be projected into a common implicit space:

$$\mathbf{H}^{(l+1)} = \sigma\left(\sum_{\tau \in \Gamma} \tilde{\mathbf{A}}_\tau \cdot \mathbf{H}_\tau^{(l)} \cdot \mathbf{W}_\tau^{(l)}\right) \quad (9)$$

where $\tau$ represents the type of nodes, $\Gamma$ is the set of node types, in our task, we have $\Gamma \in \mathbb{R}^{2 \times 1}$. $\tilde{\mathbf{A}}_\tau$ is the submatrix of $\tilde{\mathbf{A}}$, which represents the connection relationship of nodes with node type $\tau$, $\tilde{\mathbf{A}}_\tau \in \mathbb{R}^{|V| \times |v_\tau|}$. $\mathbf{W}_\tau^{(l)}$ is the transformation matrix of node type $\tau$. Different types of information are projected into a common implicit space, resulting in $\mathbf{H}^{(l+1)}$.

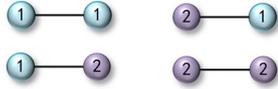

Fig. 5 Four types of meta-path. The circles represent nodes, and the lines represent edges. The colors and numbers of the circles represent the types of nodes. Nodes and edges together form a meta-path.

In addition to the differences between node types, the ways of information transfer between nodes also have differences. In heterogeneous graphs, two nodes can be connected through different semantic paths, which are called meta-paths(Shi et al., 2017). There are two types of nodes in our task, so there should be four types of meta-paths, as shown in Fig. 5. The left node in Fig. 5 is called the start node of the meta-path, and the right node is the end node. By defining the meta-path $\Phi$, semantic information of heterogeneous graphs can be more effectively mined.

*3.2.3. Multi-level multi-head attention mechanism (MMAM)*

To fully explore the complex interaction characteristics and rich semantics of heterogeneous graphs and enhance the feature extraction capabilities of LEMSF. We introduce MMAM to capture information at different levels(Vaswani et al., 2023; Veličković et al., 2018; Wang et al., 2019).

*3.2.3.1. Node level attention*

For the central node $v_i$, the importance of its multiple neighboring nodes $v_j \in N^{\Phi_p}(v_i)$ should be different in a given meta-path. $N^{\Phi_p}(v_i)$ is the set of neighbor nodes connected $v_i$ through the meta-path $\Phi_p$. The node-level attention mechanism assigns higher weights to nodes that contribute significantly to the information aggregation process while reducing the impact of noisy nodes. This approach can also reduce the negative impact of physical information errors on the network. The calculation method is shown in Eq.(10).

$$\eta_{v_i v_j}^{\Phi_p} = \sigma\left(\mathbf{v}^T \cdot \left[\mathbf{h}_{v_i} \| \mathbf{h}_{v_j}\right]\right) \quad (10)$$

where $\mathbf{v}^T$ is the learnable attention vector, $\|$ means vector concatenation, $\sigma(\cdot)$ is nonlinear activation function, the attribute vector of $v_j$ is $\mathbf{h}_{v_j}$. After normalizing $\eta_{v_i v_j}^{\Phi_p}$, the attention score $\tilde{\eta}_{v_i v_j}^{\Phi_p}$ is obtained:

$$\tilde{\eta}_{v_i v_j}^{\Phi_p} = \frac{\exp\left(\eta_{v_i v_j}\right)}{\sum_{v_j \in N^{\Phi_p}(v_i)} \exp\left(\eta_{v_i v_j}\right)} \quad (11)$$

From Eq.(11), it can be seen that the attention score depends on the features of the node pair. However, the node-level attention scores between node pairs are not symmetrical, which means $\tilde{\eta}_{v_i v_j}^{\Phi_p} \neq \tilde{\eta}_{v_j v_i}^{\Phi_p}$. This is because $\|$ is an asymmetric operation, different nodes have different neighbor nodes, so the denominator in Eq.(11) is also different.

Heterogeneous graphs have the characteristic of being scale-free, with large variations. To address the negative impact of these characteristics on attention scores, we adopt a multi-head attention calculation method to stabilize the training process. The multi-head attention mechanism also enables the output of the attention score to contain encoded representation information from different subspaces, thereby enhancing the model's expressive power. To be more specific, if there are K attention heads, we can get K representations of the central node $v_i$ after learning. The final node embedding can be obtained by contacting these K vectors, the mathematical form is shown in Eq.(12):

$$\mathbf{h}_{v_i}^{\Phi_p} = \|_{k=1:K} \sigma\left(\sum_{v_j \in N^{\Phi_p}(v_i)} \left(\tilde{\eta}_{v_i v_j}^{\Phi_p} \cdot \mathbf{h}_{v_j}\right)\right) \quad (12)$$

where $v_i$ is the end node of $\Phi_p$. $\mathbf{h}_{v_i}^{\Phi_p}$ is the embedding of the node $v_i$ that is calculated by the multi-head attention mechanism under the meta-path $\Phi_p$.

*3.2.3.2. Semantic level attention*

There are $P$ meta-paths in a graph and $\mathbf{\Phi} = \{\Phi_1, \cdots, \Phi_p, \cdots \Phi_P\}$ is the set of them. Different meta-paths have different importance at the semantic level. To this end, we use semantic attention to learn the importance differences between meta-paths. After node-level attention calculating, we can obtain $P$ feature embeddings of meta-paths, denoted as $\{\mathbf{h}_{\Phi_1}, \cdots, \mathbf{h}_{\Phi_p}, \cdots, \mathbf{h}_{\Phi_P}\}$. For $\Phi_1$, we have $\mathbf{h}_{\Phi_1} = \{\mathbf{h}_{v_1}^{\Phi_1}, \cdots, \mathbf{h}_{v_q}^{\Phi_1}, \cdots\}$, where footnote indicates of $\mathbf{h}_{v_q}^{\Phi_1}$ the end node of $\Phi_1$. $\Phi_1(v_i) = \{v_1, \cdots, v_q, \cdots\}$ is the set of end nodes in $\Phi_1$, $|\Phi_1(v_i)|$ is the quantity of the nodes. Based on the above definition, we use a single-layer multi-layer perceptron (MLP) to calculate semantic attention score, as shown in Eq.(13):

$$\tilde{\alpha}_p = \text{softmax}\left\{\frac{1}{|\Phi_p(v_i)|}\left(\sum_{i \in \Phi_p(v_i)} \boldsymbol{\mu}^T \cdot \sigma\left(\mathbf{W} \cdot \mathbf{h}_{v_i}^{\Phi_p} + \mathbf{b}\right)\right)\right\} \quad (13)$$

where $\mathbf{W}$ is the weight matrix, $\mathbf{b}$ is the bias vector, $\boldsymbol{\mu}^T$ is the learnable semantic attention vector. Finally, we perform SoftMax to obtain the normalized semantic attention score $\tilde{\alpha}_p$. For a given node $v_i$, it is the end node of multiple meta-paths at the same time. The final embedding of the node $v_i$ after weighting by the above two attentions is $v_i$:

$$\mathbf{h}_{v_i} = \sum_{p=1}^{P} \tilde{\alpha}_p \cdot \mathbf{h}_{v_i}^{\Phi_p} \quad (14)$$

From equations (10) and (13), it can be seen that the MMAM does not use the vector dot product to calculate attention scores, which avoids using similarity as the main basis for attention scoring. Although the computational cost increases slightly, this approach is more in line with the physical meaning of SAR-ATR. The physical meaning of nodes is the local components of the target. There is no clear causal relationship between the importance of neighboring components to the central component and their similarity. Therefore, we design a single-layer MLP and optimize the attention-scoring method using the classification loss.

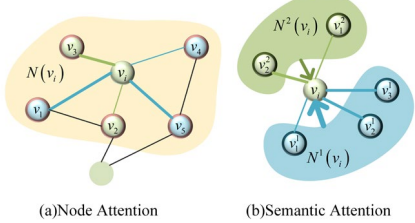

(a)Node Attention (b)Semantic Attention

Fig. 6 Schematic diagram of node attention and semantic attention. Fig (a) represents node-level attention, where the line width indicates the contribution of adjacent nodes to the central node. Fig (b) represents semantic level attention, where the green and blue arrows denote the contribution of two semantics to the central node, and the line width indicates the magnitude of the contribution.

Fig. 6 is a schematic diagram of node-level attention and semantic attention. Both of the two-layer attention mechanisms adjust the weights of the nodes during information transmission and aggregation processes. After multi-level node information aggregation, high-order semantic feature embeddings of nodes can be obtained. SAR-ATR is a graph classification task. Thus, we need to use the high-order embeddings of nodes to calculate the feature representation of the entire graph.

*3.2.3.3. Graph level attention*

From the perspective of the entire graph, the contribution of each node also varies. Therefore, the last level of MMAM is designed for the node features aggregation process of the entire graph. From the physical meaning of SAR, focusing on high-quality scattering centers (nodes in the graph) in the data has a positive impact on the final classification.

The acquisition of the feature representation of the entire graph can be seen as graph pooling, that is, based on the high-order embeddings of all nodes, pooling to obtain a vector as the representation of the graph. Therefore, the graph-level attention mechanism utilizes graph convolution to calculate attention scores. The specific process is as follows:

$$\mathbf{z} = \sigma\left(\tilde{\mathbf{A}} \cdot \mathbf{H} \cdot \mathbf{\Theta}_{att}\right) \tag{15}$$

where $\tilde{\mathbf{A}}$ is the normalized adjacency matrix, $\tilde{\mathbf{A}} \in \mathbb{R}^{|V|\times|V|}$. $\mathbf{\Theta}_{att}$ are learnable parameters, $\mathbf{\Theta}_{att} \in \mathbb{R}^{q'\times 1}$, $q'$ is the length of the nodes embeddings obtained by Eq.(14). $\mathbf{H}$ is the matrix composed of the nodes embeddings, $\mathbf{H} \in \mathbb{R}^{|V|\times q'}$, the specific mathematical form is shown in Eq.(16):

$$\mathbf{H} = \begin{bmatrix} \mathbf{h}_{v_1}(1) & \cdots & \mathbf{h}_{v_1}(q') \\ \vdots & \ddots & \vdots \\ \mathbf{h}_{v_{|V|}}(1) & \cdots & \mathbf{h}_{v_{|V|}}(q') \end{bmatrix} \tag{16}$$

The pooling approach can utilize both node characteristics and topology information. The attention score corresponding to each node is obtained after SoftMax. Based on the attention scores, we weigh and sum the node embeddings to aggregate the final feature representation of the entire graph. The mathematical form is shown in Eq.(17):

$$\mathbf{g} = \sum_{i=1}^{|V|} \mathbf{z}(i) \cdot \mathbf{h}_{v_i} \tag{17}$$

where $\mathbf{z}$ is the vector consisting of the node's attention scores, corresponding to the graph-level attention score of node $v_i$.

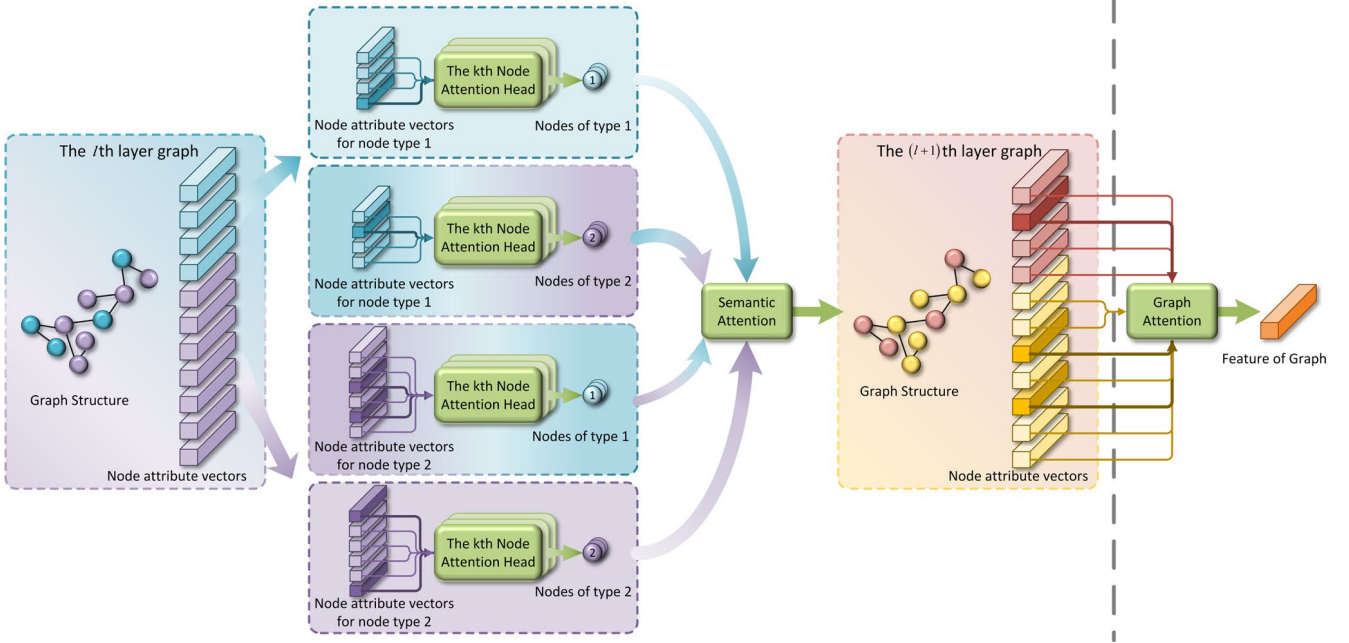

Fig. 7 Structure of LEMS. The green module in the figure represents MMAM, and the node-level attention mechanism is multi-headed. In the $l$th layer, the node types are distinguished by blue and purple, while distinguished by red and yellow in the $(l+1)$th layer. The depth of the node attribute vectors' color indicates the contribution size of the node, and the arrowed line segment indicates the direction of information transmission. The line width indicates the contribution degree of the information.

*3.2.4. The structure of LEMSF*

Summarize the above technical details and obtain a schematic diagram of the structure of LESF in Fig. 7. The left side of the dashed line shows the calculation process from the $l$th layer to the $(l+1)$th layer, and the right side of the dashed line represents the calculation process of the entire graph's feature embedding.

Due to the message-passing mechanism, the sensory field increases significantly with each additional layer in the GNN. Meanwhile, the graph constructed in our work is a complete graph. Therefore, LEMSF can achieve good results with only two convolution layers. LEMSF has a small scale and a small number of parameters, which can be trained on the CPU. This greatly improves the practical engineering deployment capability of the algorithm.

In addition, GNN has the good property of node permutation invariance. That is to say, once the node and their relationships are determined, any arbitrary node order or stretch transformation does not affect the stable extraction of features. This is consistent with the characteristic that objects with the same geometric structure have different visual performances from different perspectives in SAR images. As long as the geometric relationship construction itself is stable, regardless of how its projection shape changes, the structural information of the target can be captured anyway.

Considering the few training data of SAR, we added hyperparameters for regularization in MMAM to prevent overfitting. Specifically:

$$\begin{cases} \mathbf{h}_{v_i}^{\Phi_p} = \underset{k=1:K}{\|} \sigma \left( \sum_{v_j \in N^{\Phi_p}(v_i)} \begin{pmatrix} \alpha_{LESF} \cdot \tilde{\eta}_{v_i v_j}^{\Phi_p} \cdot \mathbf{h}_{v_j} \\ + (1 - \alpha_{LESF}) \cdot \mathbf{h}_{v_j} \end{pmatrix} \right) \\ \mathbf{h}_{v_i} = \sum_{p=1}^{P} \beta_{LESF} \cdot \tilde{\alpha}_p \cdot \mathbf{h}_{v_i}^{\Phi_p} + (1 - \beta_{LESF}) \cdot \mathbf{h}_{v_i}^{\Phi_p} \\ \mathbf{g} = \sum_{i=1}^{|V|} \gamma_{LESF} \cdot \mathbf{z}(i) \cdot \mathbf{h}_{v_i} + (1 - \gamma_{LESF}) \cdot \mathbf{h}_{v_i} \end{cases} \quad (18)$$

Comparing Eq.(12)(14)(17) with Eq.(18), it can be seen that the hyperparameters $\alpha_{LESF}$, $\beta_{LESF}$, $\gamma_{LESF}$ are added to the node, semantic, and graph attention calculation processes. In this way, we made a tradeoff between original convolution and MMAM, which can alleviate overfitting in the model.

### 3.3. Global Visual Feature Extraction (GVF) Net

#### 3.3.1. Baseline of GVF

Early target recognition models such as AlexNet and VGG improved performance by increasing network depth, which significantly raised computational requirements. ResNet, DenseNet provide better performance with smaller model sizes. However, although DenseNet has a small scale of parameters, its dense connection structure leads to a significant increase in computation. ResNet has been widely used in many fields and has excellent performance in the field of SAR-ATR (Shao et al., 2017). Hu et al. (2020) enhanced ResNet by incorporating Squeeze and Excitation Networks (SENet), which achieved better recognition results at a slight increase in the number of parameters. More importantly, the global average pooling (GAP) layer in the SENet considers all spatial information of features, which can make the model pay more attention to the global visual features of SAR images. This is in line with our philosophy in designing the GVF. As mentioned earlier, on the premise that LEMSF can effectively learn local EMS features of the target, we hope GVF to complement the global visual information that LEMSF does not focus on. Therefore, SE-ResNet18 is the baseline we chose to design the GVF.

#### 3.3.2. Network pruning

LEMSF has a very small scale of parameters and extremely fast inference speed, and we want to maintain this advantage. Thus, the main task in designing the GVF is to reduce the computational complexity and make the structure as compact as possible while maintaining the performance of the algorithm.

##### 3.3.2.1. Channel compression

Channel pruning can significantly reduce the width of the model. It has been found that the first and last residual blocks in each stage of ResNet are more sensitive to pruning than the middle block (Han et al., 2015; H. Li et al., 2017). Since each layer has different sensitivity to pruning, different pruning rates need to be determined for different layers. We evaluate the importance of each channel or layer in the model through sparse training. Specifically, we train GVF separately and introduce a scaling factor $\chi$ into each well-trained channel of the model. Then multiply $\chi$ with the output of the channel, formally:

$$x_{output} = \chi \cdot x_{conv} + \zeta \quad (19)$$

where $x_{output}$ is the output after adding the scaling factor, $x_{conv}$ is the $i$th output after the batch normalization layer. $\chi$ and $\zeta$ are learnable parameters, which represent the scaling and shift scales. L1 regularization is used to construct the loss function for training the scaling factor:

$$L_{sparsity} = loss_{classify} + \alpha_{sparity} \cdot \sum_{\chi_{ij} \in \mathrm{X}} \|\chi_{ij}\|_1 \quad (20)$$

where $loss_{classify}$ represents the loss function for training GVF. For multi-class classification problems, we use the cross-entropy as the loss function. $\chi_{ij}$ denotes the scale factor of the $j$th channel in the $i$th layer. $\alpha_{sparity}$ is the hyperparameter used to balance the two losses.

The closer the scaling factor $\chi_{ij}$ is to 0, the smaller the corresponding channel contribution is. To prevent excessive pruning from causing excessive loss of network feature capability, the global pruning ratio $\delta_{global}$ and the local pruning ratio $\delta_{local}$ is introduced to trim channels in specific convolutional layers. First, the scale factor $\bar{\chi}_i$ of each layer is obtained by averaging the scale factors of all channels in that layer. Then sort all $\bar{\chi}_i$ in ascending order. Only channels located in the top $\delta_{global}$ percent layers will be trimmed. Thus, we sort the channel scaling factor $\chi_{ij}$ in each of these selected layers in ascending order. Eventually, the channels in the top $\delta_{local}$ percent will be trimmed.

##### 3.3.2.2. Layer compression

Layer pruning can reduce the depth of the model and is implemented only on the shortcut. Layers to be trimmed are selected according to the pruning ratio $\delta_{layer}$. Before layer pruning, we first apply scaling factors and pruning ratios to construct a pruning mask based on the network structure. Note that the shortcut-connected layers should have the same number of channels. To match this, we perform logical OR operations between layers with shortcut connections and iterate according to this operation until all layers are processed. Then layer pruning is conducted on the mask based on $\delta_{layer}$ and $\bar{\chi}_i$. Finally, retrain GVF using the same training hyperparameters as normal training to fine-tune the compressed model, allowing the network to recover accuracy.

### 3.4. Fusion of Cross-Network Features

#### 3.4.1. Feature fusion subnet

To achieve high-precision recognition, it is important to design an effective fusion strategy for the different features learned by LEMSF and GVF. Since these two branches utilize different sources of information and have different focuses, the features they learn can be thought of as features of different modes. However, they both describe the same target, which means that the two features have certain semantic connections. Therefore, our main goal in designing the fusion subnet is to maintain the integrity of the specific semantics of each modality and reduce the heterogeneity between modalities, in order to achieve optimal performance.

Traditional linear fusion methods are not sufficient for modeling the complex relationships between two modalities (Kim et al., 2017). Bilinear models provide richer representations than linear fusion methods (Y. Li et al., 2017). To achieve the fusion goal, we construct an adaptive feature fusion method based on the idea of the bilinear model. The mathematical form of the bilinear model is shown below:

$$\hat{v}_i = \mathbf{v}_{LESF}^T \mathbf{W}_i \mathbf{v}_{GIF} + b_i \qquad (21)$$

where $\mathbf{v}_{LESF} \in \mathbb{R}^{N\times 1}, \mathbf{v}_{GIF} \in \mathbb{R}^{M\times 1}$ are the two feature vectors to be fused, $\mathbf{W}_i \in \mathbb{R}^{N\times M}$ are the learnable parameters, $b_i$ is the corresponding bias. $\hat{v}_i$ is the $i$th element in the fused vector $\hat{\mathbf{v}} \in \mathbb{R}^{L\times 1}$. It can be seen that if the output vector is $L$ dimensional, the number of parameters that bilinear models need to contain is $L\times(N+M+1)$. A large number of trainable parameters can limit the model structure and computational resources. To this end, we use a low-rank bilinear method proposed by (Y. Li et al., 2017) to reduce the rank of the weight matrix in Eq.(21), thus reducing the number of parameters, and Eq.(21) is transformed into:

$$\begin{aligned}\hat{v}_i &= \mathbf{v}_{LESF}^T \mathbf{U}_i \mathbf{V}_i^T \mathbf{v}_{GIF} + b_i \\ &= \mathbf{U}_i^T \mathbf{v}_{LESF} \odot \mathbf{V}_i^T \mathbf{v}_{GIF} + b_i\end{aligned} \qquad (22)$$

$\odot$ is Hadamard product. $\mathbf{W}_i = \mathbf{U}_i \mathbf{V}_i^T$ is substituted into Eq.(21), resulting in Eq.(22), where $\mathbf{U}_i \in \mathbb{R}^{N\times d}$ and $\mathbf{V}_i \in \mathbb{R}^{M\times d}$. This approach limits the rank of $\mathbf{W}_i$ to a maximum of $d$, and $d \leq \min(N,M)$. But so far, $\mathbf{U}$ and $\mathbf{V}$ still need to be set to third order. So the further improvement is shown in Eq.(23):

$$\hat{\mathbf{v}} = \mathbf{P}^T \left(\mathbf{U}^T \mathbf{v}_{LESF} \odot \mathbf{V}^T \mathbf{v}_{GIF}\right) + \mathbf{b} \qquad (23)$$

where $\mathbf{P} \in \mathbb{R}^{d\times c}$, $\mathbf{b} \in \mathbb{R}^c$, the dimension of the fused vector is $\hat{\mathbf{v}} \in \mathbb{R}^{c\times 1}$. The dimensions of $\mathbf{U} \in \mathbb{R}^{d\times N}$, $\mathbf{V} \in \mathbb{R}^{d\times M}$ and $\hat{\mathbf{v}} \in \mathbb{R}^{c\times 1}$ are decided by hyperparameters $d$ and $c$. To improve the fitting ability of the model, a nonlinear activation function is added to Eq.(23):

$$\hat{\mathbf{v}} = \mathbf{P}^T \left(\sigma(\mathbf{U}^T \mathbf{v}_{LESF}) \odot \sigma(\mathbf{V}^T \mathbf{v}_{GIF})\right) + \mathbf{b} \qquad (24)$$

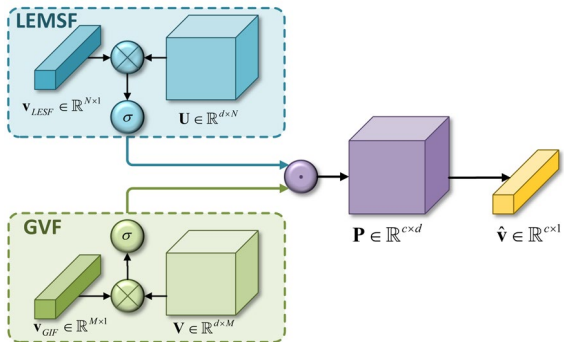

Fig. 8 Two-way feature fusion process.

It should be noted that in the Hadamard product, each input gradient directly depends on another input. Nevertheless, the output features of LEMSF and GVF belong to two different modes, and their characteristics and statistical properties may differ. To avoid interference, we first apply nonlinear activations to the two branches before performing the Hadamard product. The entire process is shown in Fig. 8. The input to the feature fusion module is the output vectors of LEMSF and GVF. Finally, a linear classifier is used to classify the fused features.

*3.4.2. Loss function construction*

The total training loss contains two parts: the classification loss and the topology loss. The cross-entropy loss function is used to calculate the classification loss:

$$L_{cls} = -\frac{1}{N}\sum_{i=1}^{N}\sum_{j=1}^{C} y_{ij}\log(\hat{y}_{ij}, \mathbf{Q}) \qquad (25)$$

where $N$ is the quantity of samples, $C$ is the number of categories, $y_{ij}$ is the value of the $i$th sample in the real label in the $j$th category. $\hat{y}_{ij}$ is the probability that the model predicts that the $i$th sample belongs to the $j$th category. $\mathbf{Q}$ is the parameter that needs to be adjusted during backpropagation.

To further improve the separability of features, we introduce the graph distance measure to construct topology loss to enhance the compactness of intra-class features (Backhausz and Szegedy, 2022; Borgs et al., 2012, 2008; Lovász, 2012). The graph distance measure defines the distance between two graphs using the norm of the difference between their adjacency matrices. It is derived from the cut norm(Frieze and Kannan, 1999), which measures the similarity of graphs from different dimensions. For two weighted graphs $G$ and $G'$, their cut distance is given by:

$$L_g(\eta,\beta) = \frac{1}{|V|}\left(\sum_{i\in V}(\eta_i - \eta_i')^2 + \sum_{i\in V}(\eta_i\eta_j\beta_{ij} - \eta_i'\eta_j'\beta_{ij}')^2\right) \qquad (26)$$

where $\eta_i$ represents the normalized node attention score, $\beta_{ij}$ is the normalized edge weight, $i,j$ is the node index. We use the distance calculated above as the second part of the loss function, resulting in the final LEMSF loss function form:

$$L_{LESF} = L_{cls} + \lambda_g L_g \qquad (27)$$

where $\lambda_g$ is the hyperparameter used to balance two types of loss functions. Under the joint supervision of classification and topology loss, the trained model will have better generalizability with limited labeled data.

## 4. Experiments

Due to the high performance of current deep learning-based algorithms on traditional evaluation methods, it is difficult to reflect the comparative effectiveness of algorithms by continuing to use these evaluation systems. We introduced a more rigorous evaluation protocol proposed by Huang et al. (2024), which is called the once-for-all (OFA) evaluation protocol. In addition, to rigorously investigate the generalizability of algorithms, we propose the less-for-more (LFM) evaluation protocol. The specific implementation plan will be explained in section 4.3.

Section 4.1 describes the dataset used in the experimental process. Section 4.3 compares the algorithms based on two more rigorous evaluation protocols, proving that LDSF has superior performance among current methods. Section 4.3.2 verifies the effectiveness of each module in LDSF through ablation studies and justifies the rationality of the algorithm structure design.

*4.1. Dataset Description*

The dataset used in the experiment is MSTAR, which has a resolution of 0.3m×0.3m and a carrier frequency of X-band. MSTAR includes 10 types of ground targets: 2S1 (self-propelled howitzer), BRDM2 (armored reconnaissance vehicle), BTR60 (armored transport vehicle), D7 (bulldozer), T62 (tank), ZIL131 (freight truck), ZSU234 (self-propelled anti-aircraft gun), T72 (tank). Each category contains a complete azimuth angle range from 0° to 360°, with a sampling interval of 5°.

Based on configuration differences between testing sets, MSTAR can be divided into four scenarios. The most common of which is the standard operating conditions (SOC) scenario, where data at 17° is used as the training set and data at 15° is used as the testing set. Specific types, quantities, and sample divisions are listed in Table 2.

**Table 2**
The serial, depression angle, and number of images available for training and testing in SOC

| Target | Train/17° | | Test/15° | |
|---|---|---|---|---|
| | Serial | Num | Serial | Num |
| 2S1 | b01 | 299 | b01 | 274 |
| BMP2 | SN_9563 | 195 | SN_9563 | 233 |
| BRDM_2 | E-71 | 298 | E-71 | 273 |
| BTR_70 | SN_C71 | 196 | SN_C71 | 233 |
| BTR_60 | k10yt7532 | 256 | k10yt7532 | 196 |
| D7 | 92v13015 | 299 | 92v13015 | 274 |
| T62 | A51 | 299 | A51 | 273 |
| T72 | SN_132 | 196 | SN_132 | 232 |
| | A64 | 298 | | |
| ZIL131 | E12 | 299 | E12 | 274 |
| ZSU_23_4 | d08 | 299 | d08 | 274 |
| Total | \ | 2934 | \ | 2536 |

The remaining three types of scenarios are known as scenarios under extended operating conditions (EOC). Due to changes in wiping angles, target settings, and other factors, the gap between the testing set and the training set becomes larger than that in SOC. This poses a more severe challenge to the performance and generalizability of algorithms. EOC-D uses data with a 30° wiping angle as the testing set, and a larger wiping angle can lead to greater differences in the visual effects of SAR images. The testing set of EOC-C is collected by configuring T72 in different ways. EOC-V testing set is obtained from different versions of BMP2 and T72 variants. The testing set under extended conditions is listed in Table 3.

**Table 3**
Test sample settings in EOC-D/C/V

| Scenario | Target | Serial | Depression | Num |
|---|---|---|---|---|
| EOC-D | 2S1 | b01 | | 288 |
| | BRDM-2 | E-71 | | 288 |
| | ZSU_23_4 | d08 | 30° | 288 |
| | T72 | A64 | | 287 |
| | Total | \ | | 1151 |
| EOC-C | T72 | S7 | | 412 |
| | | A32 | | 572 |
| | | A62 | 15°/17° | 573 |
| | | A63 | | 573 |
| | | A64 | | 573 |
| | Total | \ | | 2,703 |
| EOC-V | BMP2 | 9566 | | 428 |
| | | C21 | | 429 |
| | | SN812 | | 426 |
| | T72 | A04 | | 573 |
| | | A05 | 15°/17° | 573 |
| | | A07 | | 573 |
| | | A10 | | 567 |
| | Total | \ | | 3,569 |

### 4.2. Experimental Settings

As mentioned earlier, we first extract the EMS information based on the complex raw data of the MSTAR. The iterative termination condition for the EMSIA is set as follows: the maximum fitting percentage is 95%, the minimum peak level is -20dB and the maximum number of scattering centers is 25.

For LEMSF, which is trained at a batch size of 256 for 200 epochs. We initialize the parameters randomly and utilize the Adam optimizer to optimize our model. Moreover, the dimensionality of the semantic-level attention vectors is set to 128 and the feature dimension of the hidden layer outputs is 32. The learning rate is 0.005, and the number of attention heads is 8. The dropout rate is 0.1, and dropout is only utilized in the first convolutional layer.

In GVF, complex data are subjected to modulus operation to obtain the original SAR image. Subsequently, the derived raw images are cropped into a size of 128×128 pixels. Furthermore, gamma transformation is employed for contrast adjustment of the images.

$$V_{out} = a \cdot V_{in}^{\gamma} \quad (28)$$

where $V_{in}$ is the input raw image grayscale value, $V_{out}$ is the grayscale output value after gamma transformation, $a$ is the grayscale scaling factor, $\gamma$ is the gamma factor. In our experiment, $a=1$, $\gamma=0.6$. The initialization and training methods of GVF are the same as those of LEMSF. The learning rate of GVF is set to 0.001。

Our model is implemented based on the PyTorch framework and the Depth Graph Library (DGL). The ASC estimation algorithm is built on the MATLAB 2023b platform. The CPU and GPU used in the experiment are 13th Gen Intel(R) Core (TM) i9-13900K and NVIDIA GeForce RTX 4090 with 64GB.

### 4.3. Comparative Experiment

#### 4.3.1. Description of comparison algorithms

The comparative algorithms can be divided into three groups. The first group is classical CNN algorithms, which are all based on amplitude images. By comparing with the first group of comparison algorithms, we can examine whether using complex data can improve the accuracy of target recognition.

**ResNet18**: The issue of gradient vanishing during DNN training is effectively resolved by introducing residual blocks, which significantly enhance the performance of image recognition tasks and control the model's size. The model is initialized with the latest pre-trained weights provided by the official model during training.

**SE-ResNet18**: This model integrates the SE module into ResNet18. By dynamically adjusting the feature response between channels, it enhances the model's ability to represent features and classify images in recognition tasks. It is also the technical baseline of GVF.

**A-ConvNet**(Chen et al., 2016): This algorithm employs sparsely connected layers to reduce the number of independent trainable parameters. It is an effective recognition method under the condition of a few training SAR samples and is also a classical CNN-based method in the field of SAR-ATR. The training method and hyperparameters of the model are set according to the values stated in the official statement.

The second group of comparative algorithms extracted physical information from complex data. By comparing with this group, we can study the impact of combining physical information on recognition results. The effectiveness differences between different complex information utilization methods can be compared.

**CA-MCNN**(Y. Li et al., 2022): Based on ASC parameters, CA-MCNN reconstructed SAR images and sent them along with the original ones into the network for feature extraction. This allowed the network to capture the local features of the target.

**MS-CVNets**(Zeng et al., 2022): This algorithm designed a series of SAR target feature extraction and fusion algorithms based on complex data, which can learn target features from complex data directly.

**FEC**(Zhang et al., 2021): FEC converted ASC parameters into feature vectors through visual word packets and concatenated them with pre-trained VGG depth features to achieve the utilization of EMS parameters.

The third group of comparative algorithms all used GNN to learn the structural features of the target. However, there are differences in graph construction and GNN structure. Some of them also have utilized EMS information. Comparing with this group of algorithms is to investigate the impact of graph and network construction on the final recognition results.

**ST-Net** (Kang et al., 2023): ST-Net used the Harris operator to extract the scattering centers based on amplitude images. Then retained 9 nodes to represent the key components of the target by K-means clustering. A GNN module is constructed to learn the structural features from these 9 nodes. To maximize the restoration of ST-Net settings, we selected the top 9 nodes with the highest amplitude as nodes of the graph.

**CNN-MSGCN**(Li et al., 2023): This algorithm extracted scattering centers based on the ASC model, retaining the top 20 scattering centers in amplitude as nodes of the graph. The node attribute vectors are obtained by multiplying the CNN features of the first layer with the reconstructed component diagram and then fed into a GNN to learn structural features.

**VSFA**(Zhang et al., 2023): In addition to estimating the scattering center using the ASC model, VSFA also used SIFT to extract key nodes, and used these two types of nodes to form a graph for feature learning.

All comparison algorithm hyperparameter settings follow the original paper. All models are trained without any data augmentation.

### 4.3.2. Analysis of once-for-all evaluation (OFA) results

In this section, we evaluate algorithms using the OFA protocol (Huang et al., 2024). The OFA requires algorithms to be tested directly on different scenarios after completing one full training on the SOC training set. It places higher demands on the robustness and generalizability of algorithms. We set up three benchmarks: probability of correct classification accuracy (PCC), model size, and floating point operations (FLOPs). These benchmarks assess algorithms algorithms' performance, model size, and running efficiency respectively. The PCC value is the average of 10 experiments for each algorithm, and Table 4 contains specific information.

**Table 4**
Comparison of Algorithms Performance under OFA evaluation protocol

| Group | Model | OFA PCC (%) | | | | Model Size (MB) | FLOPs |
| --- | --- | --- | --- | --- | --- | --- | --- |
| | | SOC | EOC-D | EOC-C | EOC-V | | |
| 1 | ResNet18 | 97.85 | 61.72 | 86.66 | 86.48 | 11.15 | 3.64e9 |
| | SE-ResNet18 | 98.50 | 61.39 | 87.17 | 87.27 | 11.33 | 4.42e9 |
| | A-ConvNet(Chen et al., 2016) | 98.92 | 62.21 | 88.77 | 87.75 | **0.30** | **6.24e7** |
| 2 | CA-MCNN(Y. Li et al., 2022) | 98.81 | 68.87 | 90.61 | 90.67 | 2.83 | 9.46e8 |
| | MS-CVNets(Zeng et al., 2022) | 98.43 | 71.33 | 92.28 | 93.14 | 5.41 | 4.82e9 |
| | FEC(Zhang et al., 2021) | 98.98 | 69.68 | 91.87 | 92.32 | 16.81 | 1.61e9 |
| 3 | ST-Net(Kang et al., 2023) | 98.93 | 70.95 | 89.60 | 91.76 | 23.91 | 7.85e9 |
| | MSGCN(Li et al., 2023) | 99.06 | 74.89 | 93.34 | 90.58 | 4.54 | 1.49e9 |
| | VSFA(Zhang et al., 2023) | 99.15 | **78.96** | 95.53 | 96.11 | 1.24 | 3.06e8 |
| | LDSF(Ours) | **99.27** | 77.72 | **96.29** | **97.71** | 0.63 | 2.52e8 |

According to the results presented in Table 4, A-ConvNet has an absolute advantage in model size and computational complexity, but its PCC is the lowest. The size of LDSF comes second after A-ConvNet and outperforms other comparative algorithms significantly. Among the first group of algorithms, A-ConvNet has the best PCC compared to universal recognition methods, indicating the unnecessary to improve the SAR target recognition method according to the characteristics of SAR. The overall PCC of the second group algorithm is higher than that of the first group, proving that the phase of SAR data does indeed contain information that is beneficial for classification. The difference in PCC among the second group of algorithms illustrates that the use method of complex information will affect the performance of the algorithm. The third group of algorithms have extracted and used structural features and have better PCC in some scenarios. Especially in the EOC-D with large changes in angle, the PCC of the third group is significantly better than that of the other two groups. However, there is a clear difference in performance among the third group of algorithms, primarily due to the construction of graph and structural feature learning approaches. ST-Net limits the number of nodes when constructing a graph, and it is based on amplitude images. Therefore, the attribute vector of nodes is far less rich than that obtained using complex data.

To more easily investigate the relative relationship between algorithm results, we visualized the data in Table 4. As shown in Fig. 9. The red pentagram indicates the best algorithm in four scenarios.

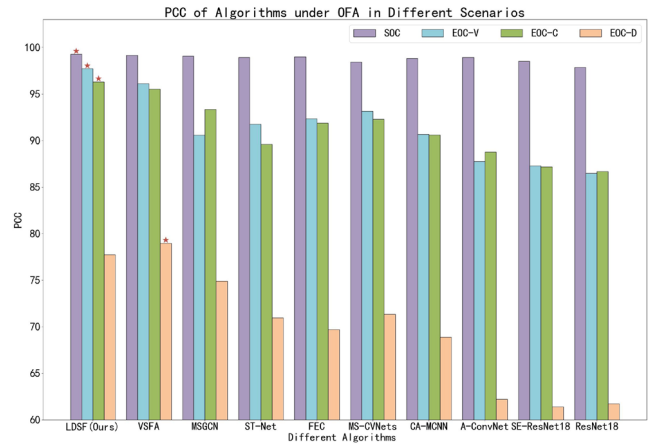

Fig. 9 Visualization of results under OFA evaluation protocol, where the red five-pointed star in the figure represents the best PCC. The difference between datasets is distinguished by different colors.

LDSF achieves the highest PCC in SOC, EOC-C, and EOC-V. However, VSFA achieves the highest PCC in EOC-D. We believe this is because VSFA uses multiple methods to extract the key nodes of the target, obtaining more nodes than other methods. Having more nodes in the graph means that the input of GNN contains more information about the target key components and structure. Thus, more information can be used for feature generalization of the pitch angle, resulting in a higher PCC for the algorithm in EOC-D. The results of ST-Net and MSGCN further support this viewpoint. Both of them have utilized homogeneous GNN networks. However, MSGCN has more than twice the quantity of nodes as ST-Net. The significant difference in quantity results in a significantly lower PCC for ST-Net with more complex structures than MSGCN. Although the quantity of nodes has a certain impact on the recognition result, a complex node extraction approach can heavily increase the computational burden. Therefore, a tradeoff needs to be made between computational complexity and recognition accuracy. As mentioned earlier, both graph construction approaches and feature learning methods can affect performance. Therefore, even though VSFA has more nodes than LDSF, the overall performance of LDSF is better. This is because LDSF had fully considered the characteristics of SAR targets and used heterogeneous graphs to better fit the real situation. Using ASC parameters as node attribute vectors maximizes the retention of EMS information. These approaches are more conducive to GNN learning effective local EMS features.

### 4.3.3. Analysis of less-for-more evaluation (LFM) results

Although the OFA protocol proposes stricter evaluation methods for algorithms, it is still based on a complete dataset. To further investigate the generalization performance of algorithms, we propose a less-for-more (LFM) evaluation protocol. MSTAR has complete azimuth angular data, but LFM prohibits algorithms from using full-angle data for training. LFM selects data within the specified azimuth angle range as the training set, while still using complete azimuth angle range data as the testing set. Depending on the limitation of the angle range, we set up three LFM test scenarios. They are training sets based on the azimuth range of -180 ° to 90 °, -90 ° to 90 °, and 0 ° to 90 °, respectively. And they are labeled as scenarios 1, 2, and 3. The specific performance of the algorithms in the above three scenarios is shown in Table 5.

**Table 5**
Comparison of Algorithms Performance under LFM evaluation protocol

| Model | OFA PCC (%) | | | | | | | | | | | |
|---|---|---|---|---|---|---|---|---|---|---|---|---|
| | SOC | | | EOC-D | | | EOC-C | | | EOC-V | | |
| | 1 | 2 | 3 | 1 | 2 | 3 | 1 | 2 | 3 | 1 | 2 | 3 |
| ResNet18 | 88.75 | 83.61 | 76.80 | 59.61 | 52.53 | 37.87 | 84.97 | 77.51 | 67.94 | 85.49 | 78.20 | 67.75 |
| SE-ResNet18 | 91.58 | 85.79 | 80.33 | 61.01 | 52.44 | 41.13 | 84.41 | 78.84 | 68.86 | 84.78 | 78.94 | 70.67 |
| A-ConvNet(Chen et al., 2016) | 92.94 | 85.54 | 79.33 | 60.58 | 53.14 | 41.51 | 86.12 | 79.48 | 70.54 | 85.20 | 76.07 | 69.32 |
| CA-MCNN(Y. Li et al., 2022) | 93.17 | 86.80 | 81.02 | 68.41 | 60.15 | 44.64 | 88.96 | 81.22 | 72.58 | 87.97 | 79.23 | 69.81 |
| MS-CVNets(Zeng et al., 2022) | 92.47 | 88.42 | 83.95 | 68.72 | 62.89 | 52.91 | 90.11 | 84.16 | 73.61 | 90.86 | 85.43 | 77.49 |
| FEC(Zhang et al., 2021) | 93.64 | 87.47 | 83.65 | 68.23 | 60.94 | 51.12 | 89.51 | 83.70 | 73.06 | 91.22 | 84.64 | 76.96 |
| ST-Net(Kang et al., 2023) | 91.89 | 89.75 | 82.68 | 68.42 | 61.09 | 52.19 | 89.03 | 81.51 | 70.69 | 89.63 | 83.73 | 76.54 |
| MSGCN(Li et al., 2023) | 93.46 | 89.61 | 84.14 | 72.77 | 69.38 | 55.95 | 92.62 | 83.74 | 74.95 | 90.58 | 83.23 | 78.08 |
| VSFA(Zhang et al., 2023) | 96.85 | 92.38 | 88.57 | 77.24 | **72.74** | 66.74 | 94.25 | **88.34** | 76.66 | 93.66 | 88.72 | 81.76 |
| LDSF(Ours) | **97.20** | **93.63** | **91.09** | **77.71** | 72.63 | **67.72** | **96.13** | 87.16 | **77.85** | **95.88** | **91.32** | **85.54** |

**Table 6**
PCC Results Obtained After Conducting Experiments on Different Modules in LDSF

| Experiment No. | LEMSF | Components of MMAM | | | GVF | PCC(%) |
|---|---|---|---|---|---|---|
| | | Node level Attention | Semantic level Attention | Graph level Attention | | |
| 1 | √ | | | | | 85.54 |
| 2 | √ | √ | | | | 92.83 |
| 3 | √ | | √ | | | 90.25 |
| 4 | √ | | | √ | | 88.17 |
| 5 | √ | √ | √ | √ | | 97.72 |
| 6 | | | | | √ | 98.19 |
| 7 | √ | | | | √ | 98.83 |
| 8 | √ | √ | √ | √ | √ | 99.27 |

We also visualized the data in Table 5 to make it easier to compare the differences between the results. Visualization results are displayed in Fig.10. It can be seen that as the range of training set angles shrinks, the PCC becomes increasingly worse. The performance of the first group algorithms decreases most significantly, which is because convolutional operations do not have rotational invariance. Therefore, when the azimuth angle is missing and the data is not enhanced by rotation, the feature extraction ability of CNN will be greatly limited. For SAR, the change in azimuth angle leads to significant differences in the backscatter performance, which further exacerbates the limitations of CNN. The performance of CA-MCNN, which only used CNN to extract visual features, also decreased significantly among the second group of algorithms. Other algorithms in the second group that utilized complex information performed slightly better, indicating that mining non-visual features from complex information helps improve the generalization ability of the algorithm. The third group of algorithms, which combined the structural features of the target, performed the best among the three groups of algorithms.

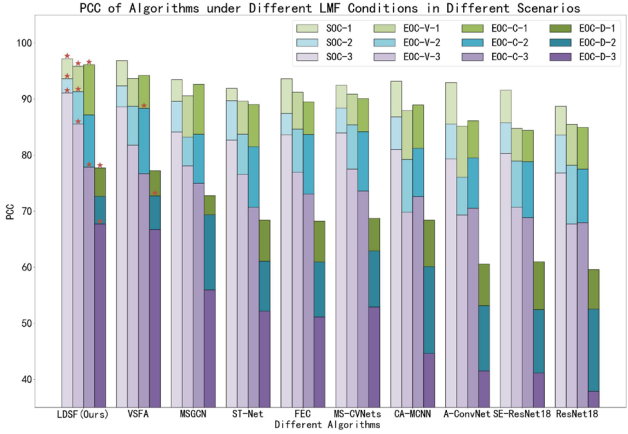

Fig. 10 Visualization of results under LFM evaluation protocol. The scene differentiation of LFM is distinguished by different colors. The difference between datasets is distinguished by the depth of colors.

The above results prove that the structural features of the target are minimally affected by the number of samples. Moreover, although some key components of SAR targets exhibit angular glint effects at different azimuth angles, GNN can learn advanced semantic information from the topological relationship between key components and the EMS of the component itself. Such ability enables the algorithm to effectively learn features closer to the essence of the target in limited angular data. In addition to achieving optimal performance in different scenarios, LDSF has minimal impact on the performance of the algorithm due to fluctuations in the quantity of training samples. This reflects the fact that LDSF has the best robustness.

*4.4. Ablation Study*

In this section, we demonstrate the effectiveness of different modules in LDSF through ablation experiments.

*4.4.1. Analysis of different modules*

*4.4.1.1. Effectiveness analysis of LEMSF and GVF*

First, we conducted ablation experiments on SOC to investigate the impact of two streams and MMAM on model performance. The specific experimental protocol and results are listed in Table 6.

Comparing experiment No.5 with the first four experiments, it can be seen that applying one part of MMAM to LEMSF can improve the performance to a certain extent. However, when the three-level mechanism is used jointly, the PCC increases significantly. This indicates that the feature information of the target is hidden in various levels of the local structure, while MMAM we designed can effectively capture this information.

Comparing experiment No.5 with experiment No.6, we found that the performance of LEMSF is slightly lower than GVF. This phenomenon can be explained from the perspective of information quantity. While LEMSF only utilizes the highlighted area of the target components, GVF utilizes the entire SAR image. The areas such as shadows, backgrounds, and outlines of the target, undoubtedly contain information beneficial for classification. Therefore, the PCC of GVF is slightly higher than LEMSF. This also indirectly indicates the necessity of building GVF, which can capture information ignored by LEMSF.

To observe more intuitively the relationship between two features of GVF and LEMSF, we visualized the two features after dimensionality reduction using t-SNE, and the results are displayed in Fig. 11. It can be seen that the features of LEMSF and GVF fall in different locations in the space, indicating that the two streams describe target characteristics from different perspectives. However, both features have good separability, which means that both streams have learned effective features.

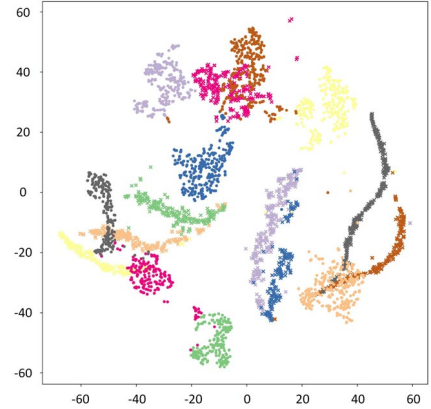

Fig. 11 Visualisation results of LEMSF and GVF features after t-SNE dimensionality reduction. Target types are distinguished by color, with features of LEMSF denoted by '×' and features of GVF denoted by '•'.

*4.4.1.2. Effectiveness analysis of loss function*

We also verified the effectiveness of the loss function designed in Section 3.4.2. We trained the model based on two loss functions, Eq.(25) and Eq.(27), and the resulting feature three-dimensional spatial distribution is shown in Fig. 12.

The features in the bottom row of Fig. 12 have better intra-class aggregation. This result suggests that adding graph measure constraints to the loss function can improve the intra-class aggregation. Comparing (a)(d) with (b)(e), we can learn that this improvement is mainly reflected in the feature learning of LEMSF.

*4.4.1.3. Effectiveness analysis of feature fusion subnet*

Table 7 lists several common feature integration methods. We compared the feature fusion subnet in LDSF with these methods to investigate the effectiveness of our method. After training the model, the model parameters are fixed, and only used different strategies for the feature fusion part. The PCC results are shown in Table 7.

**Table 7**
Ablation experiment of different components in LDSF

| | Different Fusion Methods | | | | PCC(%) |
|---|---|---|---|---|---|
| Contact | Max(·) | Min(·) | Mean(·) | Feature Fusion Subnet | |
| √ | | | | | 98.36 |
| | √ | | | | 98.32 |
| | | √ | | | 98.19 |
| | | | √ | | 97.81 |
| | | | | √ | 99.27 |

The proposed feature fusion subnet performs better than that of the traditional linear fusion strategy, as it effectively utilizes the complex relationships between the two modalities. From the results in Table 7, our method can maximize the performance of the algorithm.

The poor flexibility of physical models in some scenarios, where there may be certain model and estimation errors (Huang et al., 2024). It is with this in mind, that we have avoided a hand-designed fusion strategy. Instead, we employed an adaptive approach that utilizes classification loss to train the model, thereby balancing the contribution of physical information.

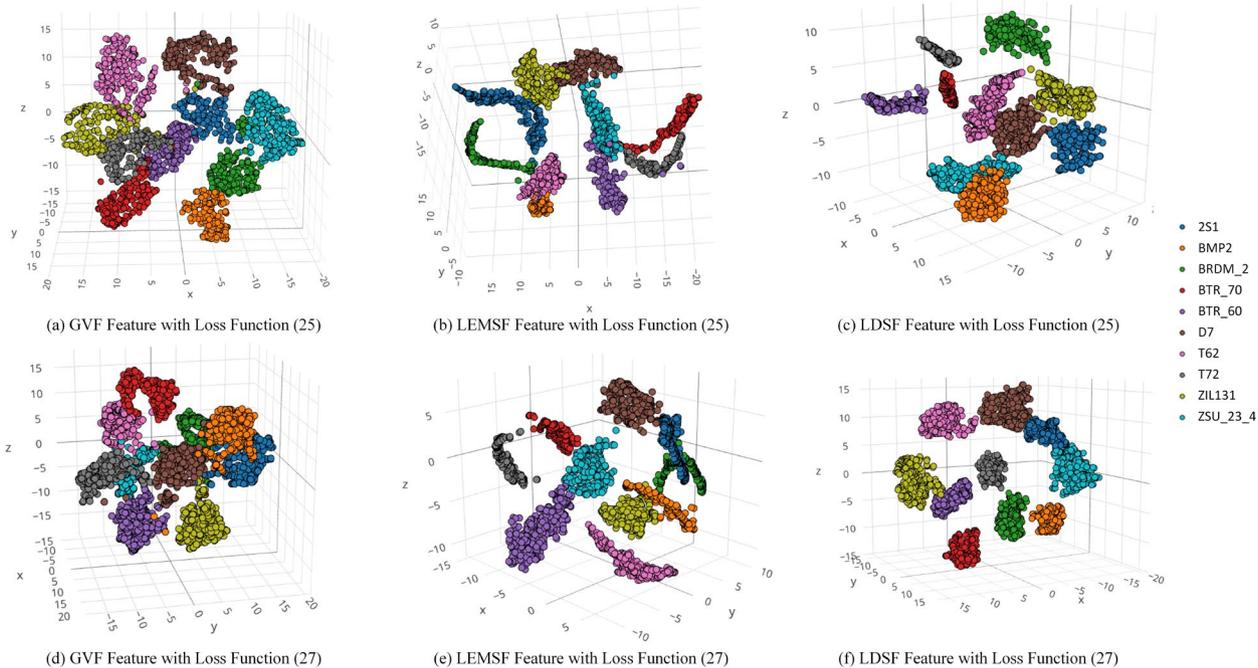

Fig. 12 Training the model with different loss functions and visualizing the features learned by GVF, LEMSF, and LDEF in 3D.

### 4.4.2. Analysis of generalization capability

In this section, we conducted experiments under the LFM protocol on LEMSF, GVF, and LDSF to investigate their generalization capabilities. The experimental results obtained in the SOC are shown in Fig. 13, where the experimental conditions for LFM are consistent with those in 4.3.3, and the fourth set of results is based on complete data sets.

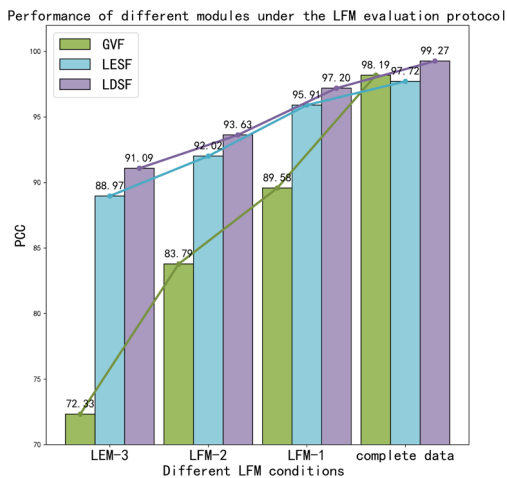

Fig. 13 Accuracy of different modules under LFM. The abscissa represents different LFM conditions, and the ordinate corresponds to the PCC results of the experiment.

As shown in Fig. 13, with the change in the quantity of training samples, GVF is the most affected. In contrast, LEMSF is significantly more stable and has a stronger generalization ability under the condition of angle missing. LDSF combining two types of features has the best performance. This further proves that effectively combining structural features can overcome the angular variability caused by SAR imaging.

## 5. Conclusion

Fusing physical knowledge can improve algorithm performance and achieve a higher level of physical interpretability. Moreover, learning from physical information can reduce the model's dependence on the data volume as well. The performance of feature fusion methods based on physical information mainly depends on the high representativeness of physical features and the effectiveness of the fusion strategy.

A novel lightweight dual-stream framework fused with physical information named LDSF is proposed in this paper. The LEMSF branch in LDSF constructed a heterogeneous graph neural network guided by a multi-level and multi-head attention mechanism to efficiently extract EMS features and topological structure features from SAR complex data. These features enable the algorithm to have a certain generalization ability and mitigate the negative impact of visual angle sensitivity on the classification. The GVF branch is responsible for capturing intuitive visual features from the image domain, thus reducing category confusion. By incorporating graph measure theory, the loss function is improved to further enhance the intra-class aggregation. The feature fusion subnet in LDSF, which takes full account of the heterogeneity and complementarities of the features of the two streams, utilizes the bilinear model to achieve an effective feature fusion. Meanwhile, under the premise of maintaining the classification performance, we control the parameter size by network pruning. This approach greatly reduces the parameter scale and computational complexity of the algorithm, making the algorithm have better engineering deployment capabilities. To investigate the actual performance of the algorithms more strictly, we conducted experiments based on two more rigorous evaluation protocols. The results show that LDSF proposed in this paper has significant advantages in terms of performance and generalization ability.

## Declaration of competing interest

We declare that we have no conflict of interest.

## Acknowledgments

This work was supported in part by the National Key Research and Development Program of China under Grant 2021YFB3100800, and in part by the Innovative Research Group Project of the National Natural Science Foundation of China under Grant 61921001.